\documentclass[mathleft]{an}

\usepackage{graphicx}
\usepackage{amsfonts,amssymb}
\usepackage{times}

\newcommand{\Rey}{\mathrm{Re}}

\newcommand{\Rin}{R_\mathrm{in}}
\newcommand{\Rout}{R_\mathrm{out}}
\newcommand{\muh}{{\mu}}
\newcommand{\etah}{\hat{\eta}}
\newcommand{\Pm}{\mathrm{Pm}}
\newcommand{\Ha}{\mathrm{Ha}}

\newcommand{\mperm}{\mu_0}
\newcommand{\mdiff}{\eta}
\newcommand{\cmnt}[1]{}
\newcommand{\comm}[1]{}
\newcommand{\ignore}[1]{}
\def\rot{\mathop{\rm rot}\nolimits}
\def\div{\mathop{\rm div}\nolimits} 
\def\gsim{\lower.4ex\hbox{$\;\buildrel >\over{\scriptstyle\sim}\;$}} 
\def\lsim{\lower.4ex\hbox{$\;\buildrel <\over{\scriptstyle\sim}\;$}} 

\def\qq{\qquad\qquad}                      
\def\q{\qquad}
\def\beg{\begin{eqnarray}}
\def\ende{\end{eqnarray}}
\renewcommand{\vec}[1]{\mbox{\boldmath $#1$}}

\newcommand{\Om}{{\it \Omega}}

\begin{document}

\sloppy

 \Pagespan{1}{}
 \Yearpublication{2009}%
 \Volume{XXX}%
 \Issue{}%

 \title{Tayler instability of toroidal magnetic fields in MHD  Taylor-Couette flows}

 \author{G. R\"udiger\thanks{Corresponding author:
   gruediger@aip.de} \and M. Schultz }

 \institute{
 Astrophysikalisches Institut Potsdam, An der Sternwarte 16, 
 D-14482 Potsdam, Germany} 
 \authorrunning{G. R\"udiger \and M. Schultz}
 \titlerunning{Tayler instability of toroidal magnetic fields} 
 \received{\today}
 \accepted{later}
 \publonline{more later}

 \abstract{The nonaxisymmetric `kink-type' Tayler instability (TI) of toroidal magnetic fields is studied for conducting incompressible fluids of uniform density between two infinitely long cylinders rotating around the same axis. The electric current  flows within  the  gap between the cylinders is axial direction. 
 For given Reynolds number of rotation 
 the   magnetic Prandtl number Pm of the liquid conductor and  the ratio of the cylinder's rotation rates are the free parameters.  It is shown   that for resting cylinders the critical Hartmann number for the unstable  modes does not depend on  Pm. By  rigid rotation  the instability is suppressed where  the critical ratio of the  rotation velocity and the Alfv\'en velocity of the field (only) slightly depends on Pm. For $\rm Pm=1$ the rotational quenching of TI takes its maximum.\\
 One also finds that rotation laws with negative shear (i.e. ${\rm d}\Om/{\rm d} R<0$) strongly  {\em destabilize} the toroidal field if the rotation is not too fast. In radiative zones of young stars, galaxies  and in the fluid crust of neutron stars this effect could have drastic implications.   For sufficiently  high Reynolds numbers of rotation the  suppression of the nonaxisymmetric magnetic instability always dominates. Superrotation laws    support the rotational stabilization  but only  for not too high Pm.\\ 
 The angular momentum transport of the instability is anticorrelated with the shear so that an eddy viscosity can be defined which proves to be positive. For  negative shear  the Maxwell stress of the perturbations remarkably contributes to the angular momentum transport.\\
We have also shown the possibility of laboratory TI experiments with a  wide-gap container filled with fluid metals  like sodium or gallium. Even  the effect of the rotational stabilization can be reproduced  in the laboratory with  electric currents of only a few  kAmp.}


 \maketitle

 \section{Motivation}
 A known  instability of toroidal fields is the current-driven (`kink-type') Tayler  instability (TI)  
 which is basically nonaxisymmetric (Tayler 1957; Vandakurov 1972; Tayler 1973; Acheson 1978). The toroidal 
 field becomes unstable against nonaxisymmetric perturbations for a sufficiently large magnetic field 
 amplitude depending on the radial profile of the field.  A global rigid rotation of the system stabilizes the TI, i.e. 
 much higher   field amplitudes can be kept stable. For the rapidly rotating 
 regime $\Om^2>\Om_{\rm A}^2$ (with $\Om_{\rm A}$ is the Alfv\'en frequency of the toroidal field) 
 the stability becomes complete, i.e. {\em all} possible modes in incompressible fluids of 
 uniform density are stable (Pitts \& Tayler 1985). 
 We shall demonstrate in the present paper how this instability and its stabilization by 
 rigid rotation  can experimentally be realized with fluid conductors like sodium and gallium. 
 There is so far no empirical or observational proof of the 
existence of the TI (Maeder \& Meynet 2005).

Another important topic in this respect is the stability of  rotation  laws 
with ${\rm d}\Om/{\rm d} R<0$ (`subrotation'). It 
is known that they  become centrifugally unstable in the hydrodynamic regime if they are steep 
enough to fulfill the Rayleigh criterion (${\rm d}(R^2 \Om)/{\rm d} R<0$). This linear instability is 
basically axisymmetric.  However,  for magnetized ideal fluids under  rapid 
rotation,  Acheson (1978) even finds instability of the nonaxisymmetric mode with $m=1$ if the shear flow is `superAlfv\'enic', i.e. 
\beg
- R \frac{{\rm d}\Om^2}{{\rm d} R} > \Om_{\rm A}^2.
\label{ach}
\ende
Hence, a nonaxisymmetric  MHD  instability  exists even  for  rather flat rotation laws if a weak toroidal magnetic field is present (despite of the rapid-rotation condition  $\Om^2>\Om_{\rm A}^2$). Of course,  relation (\ref{ach}) has no own meaning for vanishing magnetic fields. One can also say that Eq. (\ref{ach}) describes 
a {\em destabilizing} role of the differential rotation with ${\rm d}\Om/{\rm d} R<0$. The system of flow and field becomes unstable although the differential rotation alone would be  stable  and also the magnetic field alone would be  stable.

It is, of course,  important to know whether  this   result is modified for real fluids with finite values of viscosity and magnetic diffusivity. We shall show that indeed an  extreme destabilization of magnetic fields by   weak subrotation exists for moderately rapid rotation. More important, however, is the behavior of this nonaxisymmetric instability for very fast rotation as the latter tends to destroy nonaxisymmetric magnetic patterns. We shall find that it indeed  disappears for too fast rotation. The astrophysical consequences for the stability of toroidal magnetic fields in differentially  rotating  stellar radiative zones and in the fluid crust of high-spinning neutron stars might be very  strong.

Important is also the existence of a nonaxisymmetric instability for  flat subrotation laws even for current-free toroidal fields ($B_\phi\propto 1/R$) which we have called  azimuthal magnetorotational instability (AMRI, see  R\"udiger et al. 2007a). It appears   if  the shear becomes superAlfv\'enic, i.e. the magnetic Reynolds number exceeds the (high enough) Lundquist number of the toroidal field.  For   too high  Reynolds numbers, however,  also this  effect disappears. Nonuniform rotation always tends to suppress any nonaxisymmetric magnetic mode. The same phenomenon can be observed for the nonaxisymmetric modes of TI.

Another new  question arises about the role of `superrotation' (i.e. ${\rm d}\Om/{\rm d} R>0$) which  is always stable in the hydrodynamic regime.   One can  expect that  toroidal fields subject to superrotation may be stabilized. Then it should also be   true that  for  solar low latitudes, where in the bulk of the convection zone the equatorial $\Om$ increases outwards,  the toroidal field is stabilized and can be amplified to much higher values than it would be 
possible  for the opposite rotation law. Note that the sunspots with their rather high magnetic field strength appear in the  same area as the superrotation does.  An open question is whether  a rotation law with negative shear destabilizes the field so that it cannot reach high amplitudes.  In the present paper it is shown with a simplifying  cylinder geometry that indeed for not too large magnetic Prandtl numbers superrotation stabilizes toroidal magnetic fields while   subrotation strongly destabilizes toroidal magnetic fields in case that  the rotation is not too fast. The stabilization by superrotation, however, vanishes for large magnetic Prandtl number.

In the shearing-sheet box approximation  Tagger, Pellat \& Coroniti (1992) already considered nonaxisymmetric modes for   vertical fields  and also for azimuthal fields (Balbus \& Hawley 1992).
Except by the already mentioned authors, the stability problem of a system of toroidal fields and differential rotation has been studied in cylindric geometry several times (Michael 1954; Chandrasekhar 1961; Howard \& Gupta 1962; Chanmugam 1979; Knobloch 1992: Dubrulle \& Knobloch 1993; Kumar, Coleman \& Kley 1994;  Pessah \& Psaltis 2005; Shalybkov 2006) but in   all these  studies only  axisymmetric perturbations are considered. In ideal MHD also nonaxisymmetric modes have been studied for current-free toroidal fields (Ogilvie \& Pringle 1996).    Here  as a continuation of papers by R\"udiger et al. (2007a,b)  attention is focused to the nonaxisymmetric perturbation  modes with $m=1$ for real fluids. In particular, the possible  realizations of the instabilities as experiments in the (MHD-)laboratory are discussed.

\section{The Taylor-Couette geometry}
\begin{figure}
    \center \includegraphics[width=6.0cm,height=8cm]{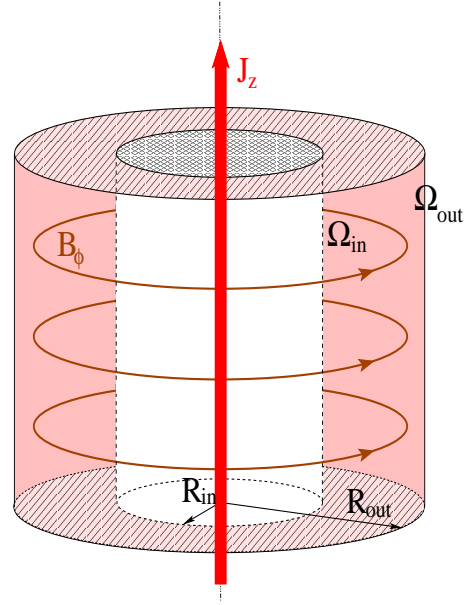}
    \caption{The  conducting fluid resides between two concentric cylinders with radii
	     $\Rin$  and  $\Rout$  rotating with $\Om_{\rm in}$ and
	     $\Om_{\rm out}$. $B_\phi$ is the  magnetic
	     field due to axial  currents inside and outside the inner
	     cylinder.}
    \label{geom}
 \end{figure}

A Taylor-Couette container is considered  confining  a  toroidal magnetic field of given amplitudes at the cylinders which rotate with different rotation rates $\Om$ (see Fig. \ref{geom}). In order to simulate the situation at the bottom of the convection zone (or even at its top) the gap between the cylinders is considered as small. For laboratory applications the case of a very wide gap is also considered. Formally, the inner radius is  $\hat\eta$\% of the outer radius. 
The extreme values of $\hat\eta=0.05$ and $\hat\eta=0.95$ are used in the present paper contrary to the calculations by R\"udiger et al. (2007a) for a medium gap of $\hat\eta=0.5$.
The fluid confined between the cylinders is assumed to be incompressible with uniform density and
 dissipative with the kinematic viscosity $\nu$ and the magnetic diffusivity
 $\mdiff$. 

Derived from  the conservation law of angular momentum the rotation law
 $\Om(R)$ in the fluid is
 \beg
    \Om(R)=a+\frac{b}{R^2}
 \ende
 with
 \beg
  a=\frac{\mu_\Om-\etah^2}{1-\etah^2} \Om_{\rm in}, \q 
  b=\frac{1-\mu_\Om}{1-\etah^2}\Rin^2 \Om_{\rm in},
 \ende
 where
\beg
\hat\eta=\frac{R_{\rm{in}}}{R_{\rm{out}}}, \q \q \q	
\mu_\Om=\frac{\Om_{\rm{out}}}{\Om_{\rm{in}}}.
\label{mu}
\ende
$\Om_{\rm in}$ and $\Om_{\rm out}$ are the imposed rotation rates of
the inner and outer cylinders with radii $R_{\rm in}$ and $R_{\rm out}$.
  After the Rayleigh stability criterion the flow is hydrodynamically stable for
 $\mu_\Om>\etah^2$. 
 We are only  interested in hydrodynamically
 stable regimes so that  $\mu_\Om>\etah^2$ must be  fulfilled. Rotation laws with ${\rm d}\Om/{\rm d}R>0$  are described by $\mu_\Om>1$ and rotation laws with ${\rm d}\Om/{\rm d}R<0$   by $\mu_\Om<1$.  $\mu_\Om=1$ gives the case of rigid rotation.
 
 Also the magnetic profiles are restricted for real fluids. The solution of the stationary induction equation without inducing shear reads 
\beg
B_\phi=A R+\frac{B}{R}.
\label{basic}
\ende
in cylinder geometry. 
 $A$ and $B$  are the fundamental
quantities;  the term $A R$ in Eq. (\ref{basic}) corresponds to  uniform axial
currents  with $I=2A$ everywhere within $R<R_{\rm out}$, and $B/R$ corresponds to
a uniform additional current only within $R<R_{\rm in}$. In the present paper we generally put $B=0$ with 
the consequence that the azimuthal magnetorotational instability (AMRI) does not appear.   The  behavior 
of the toroidal field is thus only due to TI for magnetic fields which are increasing outwards. 

It is useful to define the quantity
\beg
\mu_B=\frac{B_{\rm{out}}}{B_{\rm{in}}}
  =\frac{A R_{\rm out}+B/R_{\rm out}}
        {A R_{\rm in }+B/R_{\rm in }},
\ende
measuring the variation of $B_\phi$ across the gap. Vanishing $B$ leads to   $\mu_B=1/\hat\eta$.  For $\hat\eta\to 1$ this choice is so close to the current-free solution $\mu_B=\hat\eta$
 that the field  becomes unstable against perturbations with $m=1$ only for very high Hartmann numbers\footnote{$\rm Ha_{\rm crit}=\infty\ $  for current-free magnetic fields  without rotation}.  

In the following we  fix  $\mu_B=1/\hat\eta$ but we shall vary the magnetic Prandtl number
\beg
\rm Pm=\frac{\nu}{\eta},
\label{pm}
\ende
and also the  values of $\mu_\Om$. If the results are to be applied to  parts of  the solar convection zone  the magnetic Prandtl number must be replaced by its value of order unity for the turbulent medium.
\section{Equations and numerical model}
The dimensionless  MHD equations for incompressible fluids are
 \begin{eqnarray}
 \lefteqn{   {\rm Re}    \frac{\partial\vec{u}}{\partial t} +{\rm Re} (\vec{u} \cdot \nabla)\vec{u} =
               -\nabla P + \Delta \vec{u} + {\Ha^2}
               \rot \vec{B} \times \vec{B},} \nonumber\\
\lefteqn{   {\rm Rm}    \frac{\partial \vec{B}}{\partial t} =
                 \Delta \vec{B} + {\rm Rm}
	        \rot (\vec{u} \times \vec{B}),} 
\label{7}
\end{eqnarray}
with       $\div{\vec{u}} =  \div{\vec{B}} = 0$ and
  with the Hartmann number
 \beg
    \Ha = \frac{B_{\rm in} D}{\sqrt{\mperm \rho \nu \mdiff}}.
 \label{Hart}
 \ende
 Here  $D=\sqrt{\Rin(\Rout-\Rin)}$ is used as the unit of length,  $\eta/D$ as the 
 unit of velocity and $B_{\rm in}$ as the unit of magnetic fields. Frequencies including the  rotation $\Om$ are normalized with the inner rotation rate $\Om_{\rm in}$. The Reynolds number
 $\Rey$ is defined as 
 \beg
 \Rey=\frac{\Om_{\rm in}  D^2}{ \nu}
 \label{re}
 \ende
  and the magnetic Reynolds number as 
\beg
\rm Rm=\frac{\Om_{\rm in}D^2}{\eta}.
\label{10}
\ende 
It appears here as  useful to work with the `mixed' Reynolds number 
 \beg
\rm Rem=\sqrt{\rm Re \cdot Rm}
\label{rem}
\ende
which is symmetric in $\nu$ and $\eta$ as it is also the Hartmann number.  
For ${\rm Pm}=1$ it is ${\rm Re}={\rm Rm}={\rm Rem}$. It is sometimes useful to use the Lundquist number ${\rm S}= \sqrt{\rm Pm}\ {\rm Ha}$. 
The ratio of Rem and Ha, 
\beg
{\rm Mm}= \frac{{\rm Rem}}{{\rm Ha}},
\label{Mach}
\ende
is called the magnetic Mach number.

Applying the usual normal mode analysis, we look for solutions of the
linearized equations of the form
\beg
F=F(R){\rm{exp}}\bigl({\rm{i}}(kz+m\phi+\omega t)\bigr).
\label{nmode}
\ende
Using Eq. (\ref{nmode}), linearizing the Eq. (\ref{7})
and representing the result as a system of first order equations, one finds
\begin{eqnarray}
\lefteqn{\frac{{\rm d}u_R}{{\rm d}R}+\frac{u_R}{R}+{\textrm i}\frac{m}{R}u_\phi+{\textrm i}ku_z=0,}
\nonumber \\
\lefteqn{\frac{{\rm d}P}{{\rm d}R}+{\textrm i}\frac{m}{R}\phi_u+{\textrm i}kZ
+\left(k^2+\frac{m^2}{R^2}\right)u_R+}
\nonumber \\
&& \qq \q +{\textrm{iRe}}(\omega+m\Om)u_R-2\Om {\textrm{Re}} u_\phi- 
\nonumber \\
&& \qq \q -{\textrm{iHa}}^2 mA b_R +2{\textrm{Ha}}^2 A
b_\phi=0,
\nonumber \\
\lefteqn{\frac{{\rm d}\phi_u}{{\rm d}R}-\left(k^2+\frac{m^2}{R^2}\right)u_\phi-
{\textrm{iRe}}(\omega+m\Om)u_\phi+}
\nonumber \\
&& \qq \q +2{\textrm i}\frac{m}{R^2}u_R
-\frac{{\textrm{Re}}}{R}\frac{{\rm d}}{{\rm d}R}\left(R^2\Om\right)u_R+
\nonumber \\
&&\qq \q +2{\textrm{Ha}}^2A b_R
+{\textrm{iHa}}^2 mA b_\phi
-{\textrm{i}}\frac{m}{R}P=0,
\nonumber \\
\lefteqn{\frac{{\rm d}Z}{{\rm d}R}+\frac{Z}{R}-\left(k^2+\frac{m^2}{R^2}\right)u_z-
{\textrm{iRe}}(\omega+m\Om)u_z-}
\nonumber \\
&& \qq \q -{\textrm i}kP+{\textrm{iHa}}^2mA b_z=0,
\nonumber \\
\lefteqn{\frac{{\rm d}b_R}{{\rm d}R}+\frac{b_R}{R}+{\textrm i}\frac{m}{R}b_\phi+{\textrm i}kb_z=0,}
\nonumber \\
\lefteqn{\frac{{\rm d}b_z}{{\rm d}R}-\frac{{\textrm i}}{k}\left(k^2+\frac{m^2}{R^2}\right)b_R
+{\textrm{PmRe}}\frac{1}{k}(\omega+m\Om)b_R+}
\nonumber \\
&& \qq \q +\frac{1}{k}\frac{m}{R}\phi_B-\frac{1}{k}mA u_R=0,
\nonumber \\
\lefteqn{\frac{{\rm d}\phi_B}{{\rm d}R}-\left(k^2+\frac{m^2}{R^2}\right)b_\phi
-{\textrm{iPmRe}}(\omega+m\Om)b_\phi+}
\nonumber \\
&&  + {\textrm i}\frac{2m}{R^2}b_R
-R u_R
+{\textrm{PmRe}}R\frac{{\rm d}\Om}{{\rm d}R}b_R
 +{\textrm i}mA u_\phi =0,
\label{syst}
\end{eqnarray}
where $\phi_u$, $Z$ and $\phi_B$ are defined as
\begin{equation}
\phi_u=\frac{{\rm d}u_\phi}{{\rm d}R}+\frac{u_\phi}{R},
\ \ \ \
Z=\frac{{\rm d}u_z}{{\rm d}R}, \ \ \ \ 
\phi_B=\frac{{\rm d}b_\phi}{{\rm d}R}+\frac{b_\phi}{R}
\label{def}
\end{equation}
and $A=1/R_{\rm in}$ ($R_{\rm in}$ in units of $D$).

 An appropriate set of ten boundary conditions is needed to solve the
system (\ref{syst}).  For the velocity the boundary conditions are always
no-slip,
\beg
u_R=u_\phi=u_z=0.
\label{ubnd}
\ende
For conducting walls the radial component
of the field and the tangential components of the current must vanish,
yielding
\beg
{\rm d}b_\phi/{\rm d}R + b_\phi/R = b_R = 0.
\label{bcond}
\ende
These boundary conditions are applied at both $R_{\rm{in}}$ and $R_{\rm{out}}$. The wave number 
is varied as long as  for given Hartmann number the Reynolds number takes its minimum. 
The procedure is already  described in 
detail by Shalybkov, R\"udiger \& Schultz (2002). One immediately finds that the sign of the real wave 
number $k$ is free so that with the solution for $k$ also another one  with $-k$ exists. For containers bounded in $z$  standing waves can thus  develop.

For insulating walls the boundary conditions are  more complicated (see R\"udiger et al. 2007b).

 The necessary and sufficient condition for the stability of toroidal fields in 
ideal Taylor-Couette flows  against 
axisymmetric perturbations is by  Michael (1954) and reads  
\beg
\frac{1}{R^3}\frac{{\rm{d}}}{{\rm{d}}R}(R^2\Omega)^2 - \frac{R}{\mu_0 \rho}
\frac{{\rm{d}}}{{\rm{d}}R}\left( \frac{B_\phi}{R} \right)^2 > 0.
\label{mich}
\ende
 Fields which are not steeper than $B_\phi\propto R$  are thus always stable against $m=0$ perturbations if  the rotation law is also stable.
Tayler (1973) found the necessary and sufficient condition
\beg
\frac{{\rm d}}{{\rm d}R}\left(RB_\phi^2\right) < 0
\label{ddR}
\ende
for   stability of an ideal nonrotating fluid against nonaxisymmetric disturbances. Our field profile $B_\phi=AR$  is thus stable against $m=0$ and unstable for sufficiently large field amplitudes, i.e. ${\rm Ha}>{\rm Ha}_{\rm crit}$. The same is true for nearly uniform fields with $\mu_B \simeq 1$. A general  criterion for rotating flows does not exist.

First   the system (\ref{syst}) is used to demonstrate that for resting  cylinders the ${\rm Ha}_{\rm crit}$ does not depend on the magnetic Prandtl number Pm. To this end in the equations all terms resulting from the rotational influence are canceled. The frequency $\omega$ is replaced by $\omega/{\rm Re}$. Furthermore,  the flow components
$u_\phi$ and $u_z$  are replaced by $-{\rm i}u_\phi$  and  $-{\rm i}u_z$ and the field component $b_R$  is replaced by ${\rm i}b_R$. It results
\begin{eqnarray}
\lefteqn{\frac{{\rm d}u_R}{{\rm d}R}+\frac{u_R}{R}+\frac{m}{R}u_\phi+ku_z=0,}
\nonumber \\
\lefteqn{\frac{{\rm d}P}{{\rm d}R}+\frac{m}{R}\phi_u+kZ
+\left(k^2+\frac{m^2}{R^2}\right)u_R+}
\nonumber \\
&& \qq \q +{\textrm{i}}\omega u_R+ 
\frac{{\textrm{Ha}}^2 m}{R_{\rm in}} b_R +\frac{2{\textrm{Ha}}^2} {R_{\rm in}}
b_\phi=0,
\nonumber \\
\lefteqn{\frac{{\rm d}\phi_u}{{\rm d}R}-\left(k^2+\frac{m^2}{R^2}\right)u_\phi-
{\textrm{i}}\omega u_\phi
-2\frac{m}{R^2}u_R-}\nonumber\\
&& \qq \q -\frac{2{\textrm{Ha}}^2} {R_{\rm in}} b_R
-\frac{{\textrm{Ha}}^2  m}{R_{\rm in}} b_\phi
+\frac{m}{R}P=0,
\nonumber \\
\lefteqn{\frac{{\rm d}Z}{{\rm d}R}+\frac{Z}{R}-\left(k^2+\frac{m^2}{R^2}\right)u_z-
{\textrm{i}}\omega u_z+}
\nonumber \\
&& \qq \q +kP-\frac{{\textrm{Ha}}^2 m}{R_{\rm in}} b_z=0,
\nonumber \\
\lefteqn{\frac{{\rm d}b_R}{{\rm d}R}+\frac{b_R}{R}+\frac{m}{R}b_\phi+kb_z=0,}
\nonumber \\
\lefteqn{\frac{{\rm d}b_z}{{\rm d}R}+\frac{1}{k}\left(k^2+\frac{m^2}{R^2}\right)b_R
+{\textrm{Pm}}\frac{{\rm i}}{k}\omega b_R+}
\nonumber \\
&& \qq \q +\frac{m}{kR} \phi_B-\frac{m}{k R_{\rm in}} u_R=0,
\nonumber \\
\lefteqn{\frac{{\rm d}\phi_B}{{\rm d}R}-\left(k^2+\frac{m^2}{R^2}\right)b_\phi
-{\textrm{iPm}}\omega b_\phi-}
\nonumber \\
&& \qq \q - \frac{2m}{R^2}b_R
-R u_R
 +\frac{m}{R_{\rm in}} u_\phi =0,
\label{app}
\end{eqnarray}
Note that the  magnetic Prandtl number only survives together with ${\rm i}\omega$ which is purely imaginary for marginal instability. Hence, in the real part of the system (\ref{app}) the frequency terms including the Pm do not appear. The only free parameter in the real part of the system (\ref{app}), therefore,  is the critical Hartmann number which results  thus as equal for all Pm. 
Shalybkov (2006) has given a similar result but only for $m=0$.
\begin{figure}[h]
   \includegraphics[width=8.5cm,height=6.0cm]{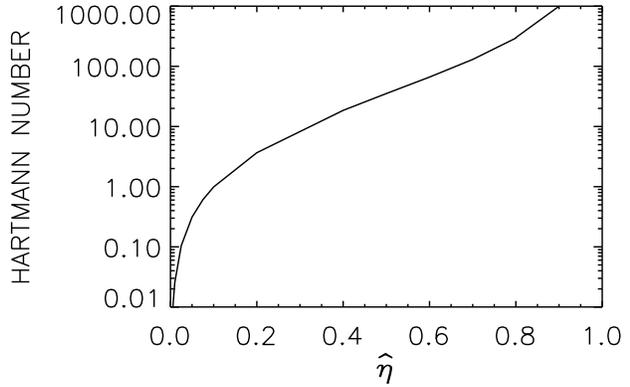}
  \caption{The critical Hartmann number for $\Om=0$ for various container gaps  ($\hat\eta$) between conducting cylinders. The numbers are valid for all magnetic Prandtl numbers, see text. }  
    \label{Ha1}
 \end{figure}

The  Hartmann numbers ${\rm Ha}_{\rm crit}$ for various $\hat\eta$ are given in Fig. \ref{Ha1}. They vary over many orders of magnitude and become very small for wide gaps ($\hat\eta\to 0$). For small $\hat\eta$  the critical Hartmann number vanishes like
\beg
{\rm Ha}_{\rm crit} \propto {\hat\eta}^{1.5}
\label{Heta}
\ende
with a  factor of about 25. Our calculation with the smallest  $R_{\rm in}$ concern  $\hat\eta=0.001$  and lead to $\rm Ha= 0.00075$.

In the present paper the gap between the cylinders is assumed as narrow ($\hat\eta=0.95$) and in another computation as wide ($\hat\eta=0.05$) so that the unit of distances, $D$, is  the same in both cases for fixed outer radius. The narrow gap model serves to astrophysical discussions (solar tachocline, neutron star crust, supergranulation layer) while  the wide gap results are needed to prepare future laboratory experiments.   We have shown with similar models that indeed  wide gaps are much more suitable for TI  experiments with liquid metals like sodium or gallium than narrow gaps  (R\"udiger et al. 2007b).
\section{Narrow gap}
For $\hat\eta=0.95$ it is $\mu_B=1.05$.  The critical Hartmann number for $\rm Re=0$ is 3061 for all Pm (see Fig. \ref{Ha1}). The Rayleigh limit for centrifugal instability is $\mu_\Om=\hat\eta^2=0.9025$.
\subsection{Rigid rotation}
We start with the simplest case of the interaction of toroidal field and global rotation, i.e. with the marginal instability for  rigid rotation  ($\mu_\Om=1$). 
 
\begin{figure}[htb]
   \includegraphics[width=8.5cm]{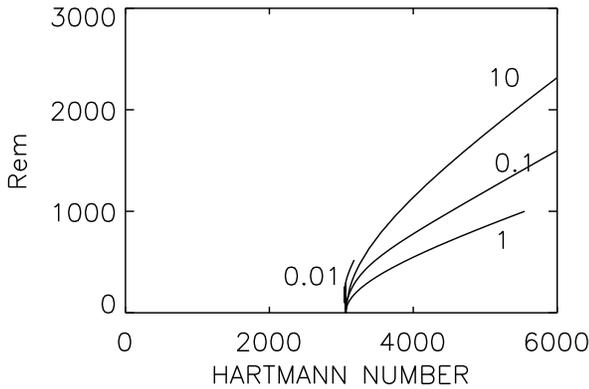}
 \caption{Narrow gap ($\hat\eta=0.95$): The suppression of the TI by rigid rotation. The curves are marked with their magnetic Prandtl number Pm. The rotational suppression of TI is weaker  for  $\rm Pm \neq 1$ than for $\rm Pm=1$. Rotating fluids with $\rm Pm=1$ allow the strongest fields to be stable.}
    \label{fig5a}
 \end{figure}
 
The results of the calculations are shown in Fig.~\ref{fig5a} in the  plane Rem-Ha. As we already know the 
critical Hartmann number Ha for $\rm Rem =0$ does not depend on the magnetic Prandtl number. The new result 
is  that the critical  Ha always grows for growing rotation rate. This is the {\em stabilizing} action  of 
rotation. In the representation of Fig.~\ref{fig5a} (where the parameters on both   axes are symmetric 
in $\nu$ and $\eta$) the growth of the critical magnetic amplitudes  is strongest for ${\rm Pm}=1$ but  it 
becomes weaker for ${\rm Pm}\neq 1$.  For ${\rm Rem}\lsim 500$ the differences of the critical magnetic 
fields for  Pm varying over three orders of magnitude are surprisingly small.  

For higher values of 
Rem the curves seem to represent a {\em linear} relation between Rem and Ha. 
 Note that  in Fig. \ref{fig5a} the magnetic Mach  number
\beg
{\rm Mm} = \frac{\Om D}{B_{\rm in}/\sqrt{\mu_0\rho}}
\label{HaRem}
\ende
 is the smallest for ${\rm Pm}=1$. If this behavior  remained true 
also for much higher Ha 
then the consequences should be strong:  In turbulent fluids and/or in simulations with ${\rm Pm}=1$  rather  
strong magnetic fields remain stable which for the real $\Pm\neq 1$ are already unstable.  For 
small Pm (stellar radiative interior) and high Pm  (galaxies, neutron stars) the magnetic instability is  much more 
efficient and already works for much smaller magnetic field strengths.

  For neutron stars the ratio (\ref{HaRem}) yields
\beg
{\rm Mm}\simeq \frac{3\cdot 10^{14}\ {\rm Gauss}}{B_\phi},
\label{HaRem1}
\ende
so that $B_\phi\simeq 3\cdot 10^{14}$ Gauss is the critical value for  the toroidal field. We have here used the numerical values $\Om=100$ s$^{-1}$, $\rho\simeq 10^{13}$ g/cm$^3$ and $H\simeq 3\cdot 10^5$ cm.

\begin{figure}[htb]
   \includegraphics[width=8.5cm]{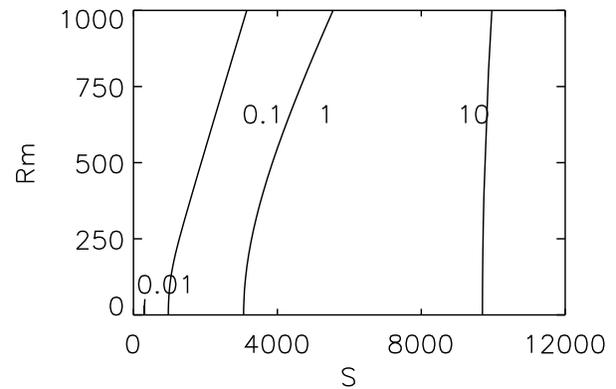}
 \caption{The same as in Fig. \ref{fig5a} but for fixed magnetic diffusivity $\eta$ and varied viscosity.  High viscosity stabilizes the magnetic fields and reduces the rotational influence.}
    \label{fig5b}
 \end{figure}

The same data are plotted in Fig.~\ref{fig5b} in the Rm-S plane. Again the curves are marked with their magnetic Prandtl numbers. One can read this plot in the sense that rotation, magnetic field  and magnetic diffusivity $\eta$ are given and the viscosity (in units of $\eta$) is varied.   For high viscosity the fields must be much stronger to become unstable. One finds a distinct stabilizing  influence of the viscosity. The rotational influence, however, is  weaker for high viscosity.

The results are applied to the bottom of the convection zone. The theory of the advection-dominated  dynamo requires the  small value of $10^{11}$ cm$^2$/s for the eddy diffusivity $\eta$ while the eddy viscosity $\nu$ should be larger for the explanation of the differential rotation pattern so that ${\rm Pm}\geq 10$. With $\Om \simeq 2\cdot 10^{-6}$ s$^{-1}$ and $D\simeq 10^{10}$ cm we find ${\rm Rm}\simeq 2000$. Figure~\ref{fig5b} yields $\rm S\leq 10^4$ for stability. With $\rho\simeq 0.1$ g/cm$^3$ the upper limit for stable toroidal fields becomes   100 kGauss. Stronger fields will not be stable against the Tayler instability. This value does not grow if the non-uniformity of the rotation (i.e. superrotation) is included due to the high value of Pm (see below).

\subsection{Nonuniform rotation} 
\begin{figure}[htb]
   \includegraphics[width=8.5cm]{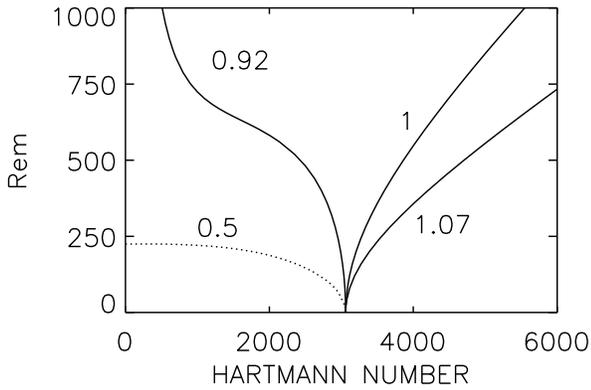}
 \caption{Reynolds number vs.  Hartmann number of the TI for subrotation ($ \mu_\Om=0.5$, $ \mu_\Om=0.92$), rigid rotation ($\mu_\Om=1$) 
           and superrotation ($\mu_\Om=1.07$).  Note the rotational quenching  for $\mu_\Om=1$, the strong destabilization by subrotation and the stabilization by superrotation. Conducting cylinder walls, $\hat\eta=0.95$, $\rm Pm=1$.}
    \label{f3}
 \end{figure}

The form of the rotation law is now changed  for various magnetic Prandtl numbers. Rotation laws with negative shear (here $\mu_\Om=0.5$ and $\mu_\Om=0.92$) and superrotation laws (here $\mu_\Om=1.07$) are investigated. The rotation law with $\mu_\Om=0.5$  is centrifugally unstable also without  magnetic field. It is given only for demonstration. The remaining  rotation laws are  stable in the hydrodynamic regime.  One finds   the numerical results for marginal stability of the $m=1$ mode  in Fig. \ref{f3}.  Superrotation  stabilizes the field more than  solid-body rotation. For subrotation the behavior is opposite.  While for rigid rotation and superrotation the critical Hartmann numbers grow for growing Reynolds number Rem,     for subrotation the Ha  become smaller so that finally the shear becomes superAlfv\'enic. This effect is in accordance with the Acheson relation   (\ref{ach}).
 
 \begin{figure}
  \vbox{
  \includegraphics[width=8.5cm]{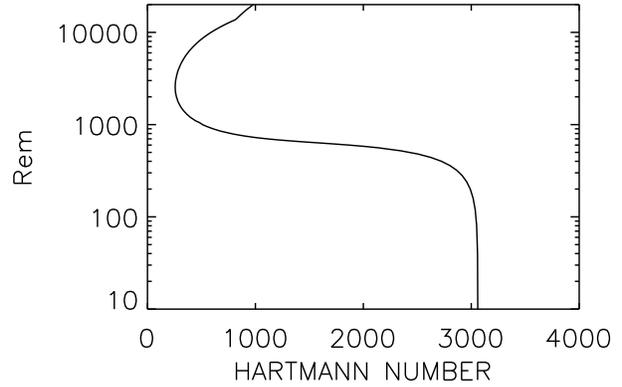}
  }
  \caption{The same as in Fig. \ref{f3} for  $ \mu_\Om=0.92$ but for much higher Reynolds numbers. The rotational destabilization 
           does only hold for medium Reynolds numbers. For faster rotation the rotational stabilization  dominates. $\rm Pm=1$.}
  \label{f6}
\end{figure}
This finding, however, cannot be the final answer. Very intensive  differential rotation always tends to suppress   nonaxisymmetric magnetic patterns. We  therefore expect that  for  higher Reynolds numbers  the Eq. (\ref{ach})
  looses its meaning. In Fig. \ref{f6} the marginal-instability curve for  $\mu_\Om=0.92$  is thus followed to very  high values of the  Reynolds number. The result of the calculations is that also for subrotation the basic rotation finally suppresses the instability   if $\rm Rem \gsim 3000$. 
  
  Up to this value the differential rotation acts  {\em destabilizing}. In regions of convection zones with ${\rm d} \Om/{\rm d} R<0$  much weaker toroidal fields become unstable than for  ${\rm d} \Om/{\rm d} R>0$.
This  finding should have strong implications for the electrodynamics of rotating stars. Young stars typically rotate ten times faster then old stars such as the Sun.  For the supergranulation part of the solar convection zone  the magnetic Reynolds number  does not exceed (say) 300. For younger solar-type stars, however, it can easily reach the  value of 3000. For all Pm and a   Reynolds number   of order 3000 the field becomes unstable already for very small Ha of order 100. Hence, the maximum stable toroidal fields in this zone for which  the helioseismology provides a clear subrotation is much weaker for fast rotating stars than for the slow-rotating Sun. Fast rotators are thus not able to accumulate strong toroidal field which could  be observed as starspots close to the equator.

In Fig.~\ref{f3} also the (dashed) curve for $\muh_\Om=0.5$ is given for comparison. Such a  rotation law is unstable  without magnetic field for $m \geq 0$. The magnetic  field additionally destabilizes  the  rotation law so that for $\rm Ha=3061$ the TI  works even without any  rotation. 

\begin{figure}[htb]
\hspace{-1.3cm}  
  \includegraphics[width=10.5cm,height=8cm]{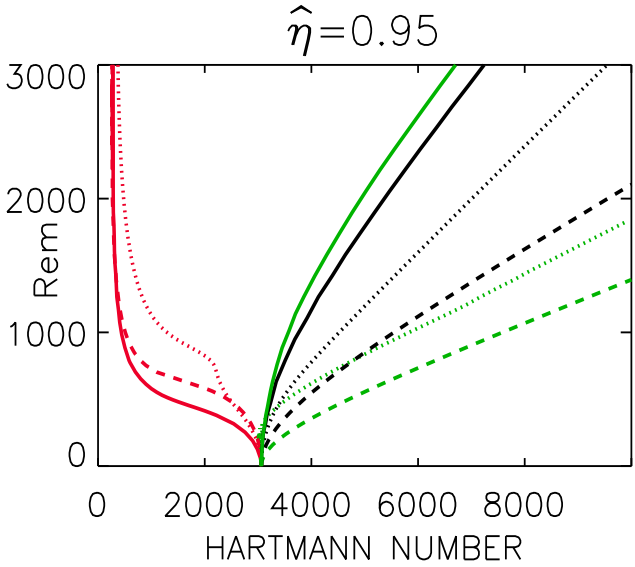}
     \caption{The same as in Fig.  \ref{f3} but for  $\rm Pm=0.1$ (dotted), $\rm Pm=1$ (dashed) 
     and $\rm Pm=10$ (solid). The curves for negative shear ($\mu_\Om=0.92$, subrotation) are red and the curves for positive shear ($\mu_\Om=1.07$, superrotation) are green. For large Pm superrotation yields  less stability  than rigid rotation.}
    \label{f8}
 \end{figure}

The differences of the results for  subrotation and for superrotation  only appear for  faster rotation  but it is until now unclear  how strong the magnetic  fields  must be.
 The calculations for nonuniform rotation laws must   be extended to  smaller and higher 
 magnetic Prandtl numbers. Figure~\ref{f8} gives the  results for $\rm Pm=0.1$ and 
  $\rm Pm=10$. They can be best written  with the characteristic numbers Rem and Ha 
 because in this formulation the differences for the small and the large magnetic Prandtl number are 
 smallest. For small magnetic Prandtl number the stabilizing influence of superrotation is   
 stronger than for high magnetic Prandtl numbers. Note that  for $\rm Pm=10$ the 
 stabilization of the magnetic field  by superrotation is even smaller than that of rigid rotation. It is not clear whether for even larger magnetic Prandtl numbers rotation laws with positive shear become able to destabilize the magnetic fields. 
 
 Written with Rem and Ha 
 one finds for given rotation law with $\mu_\Om \simeq 1$ that $\rm Pm=1$  yields the most effective stabilization of the toroidal magnetic field. For $\rm Pm \neq 1$ the stabilization is stronger for small Pm than for large Pm. 
 The differences, however,  are small. For subrotation ($\mu_\Om <1$) the very effective rotational destabilization of the magnetic field  hardly depends on the magnetic Prandtl number.

In summary, we have shown for the container with the narrow gap that without  rotation   the critical Hartmann number is $\rm Ha=  3061$ independent 
of the magnetic Prandtl number. This value is increased for solid-body rotation and for superrotation but 
it is drastically reduced to about 100  for subrotation with ${\rm Rem}\geq 1000$. There is no strong 
dependence of these characteristic numbers  on the magnetic Prandtl number. However, for too fast rotation (${\rm Rem}> 3500$) the   destabilization changes to rotational  stabilization  (see Fig. \ref{f6}). On the other hand, the role of superrotation to stabilize the magnetic field changes to destabilization for too high magnetic Prandtl number.
\subsection{The angular momentum transport}

The solutions of the linear equations are free of an arbitrary real parameter of any sign. We do  not know, therefore, the sign of the flow and/or the field. However, for quadratic expressions such as the correlation tensor or the electromotive force one can find  the signs as all the solutions are multiplied with one and the same parameter.

Let us apply this idea to the angular momentum transport
\begin{equation}
T_R=\langle u_R' u_\phi' - \frac{1}{\mu_0\, \rho} B_R' B_\phi'\rangle .
\label{T}
\end{equation}
The average procedure consists of  an integration  over the azimuth $\phi$. The question  we  shall answer is whether $T_R$ and ${\rm d}\Om/{\rm d}R$ are anticorrelated. If this is true then the angular momentum flows towards the minimum of the angular velocity, and one can introduce an eddy viscosity $\nu_{\rm T}$ in accordance to
\begin{equation}
T_R=- \nu_{\rm T} R \frac{{\rm d}\Om}{{\rm d}R}
\label{T1}
\end{equation}
with  positive $\nu_{\rm T}$.  

\begin{figure}[htb]
   \vbox{ 
   \includegraphics[width=8.5cm,height=4.5cm]{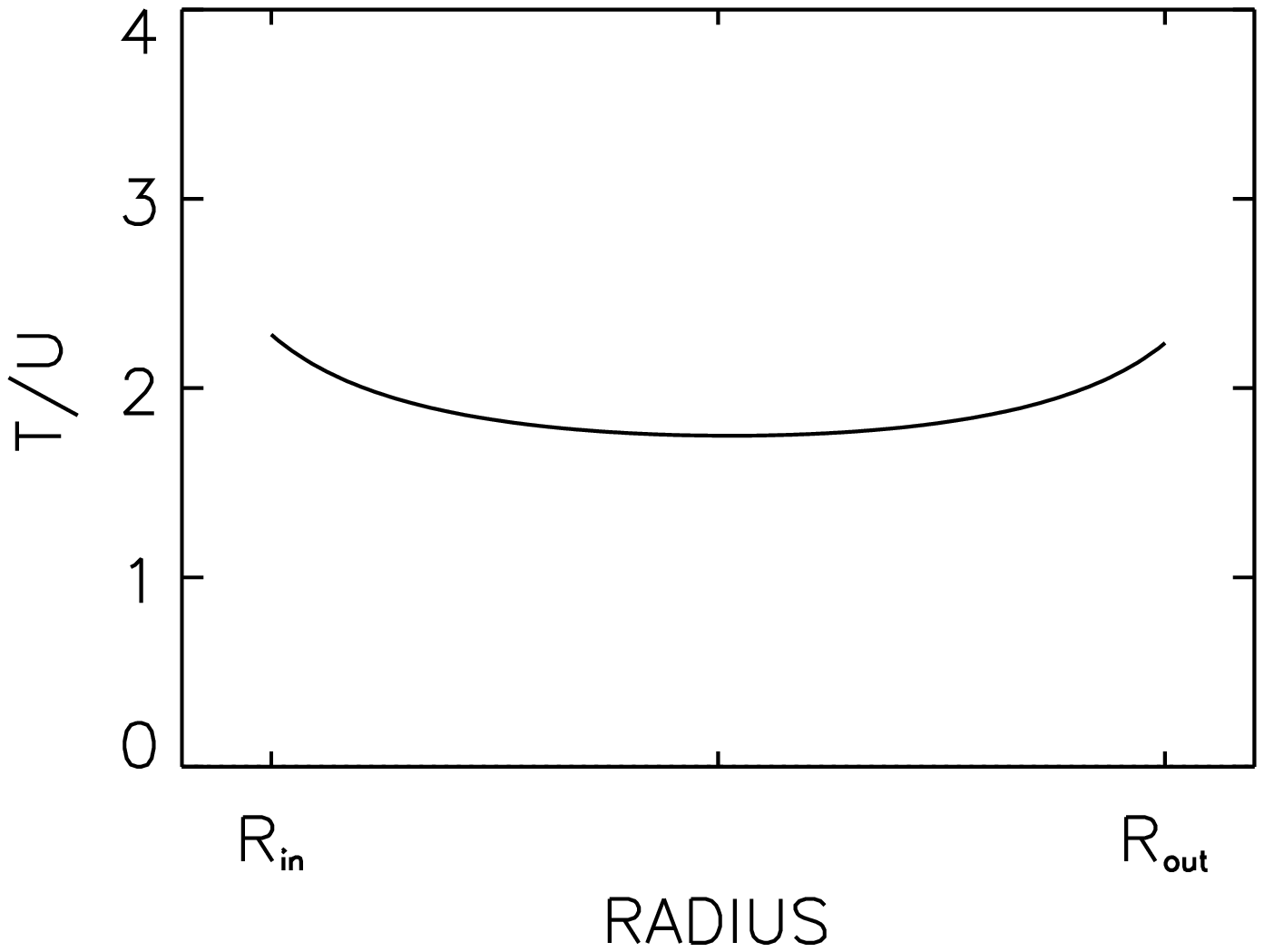}
    \includegraphics[width=8.5cm,height=4.5cm]{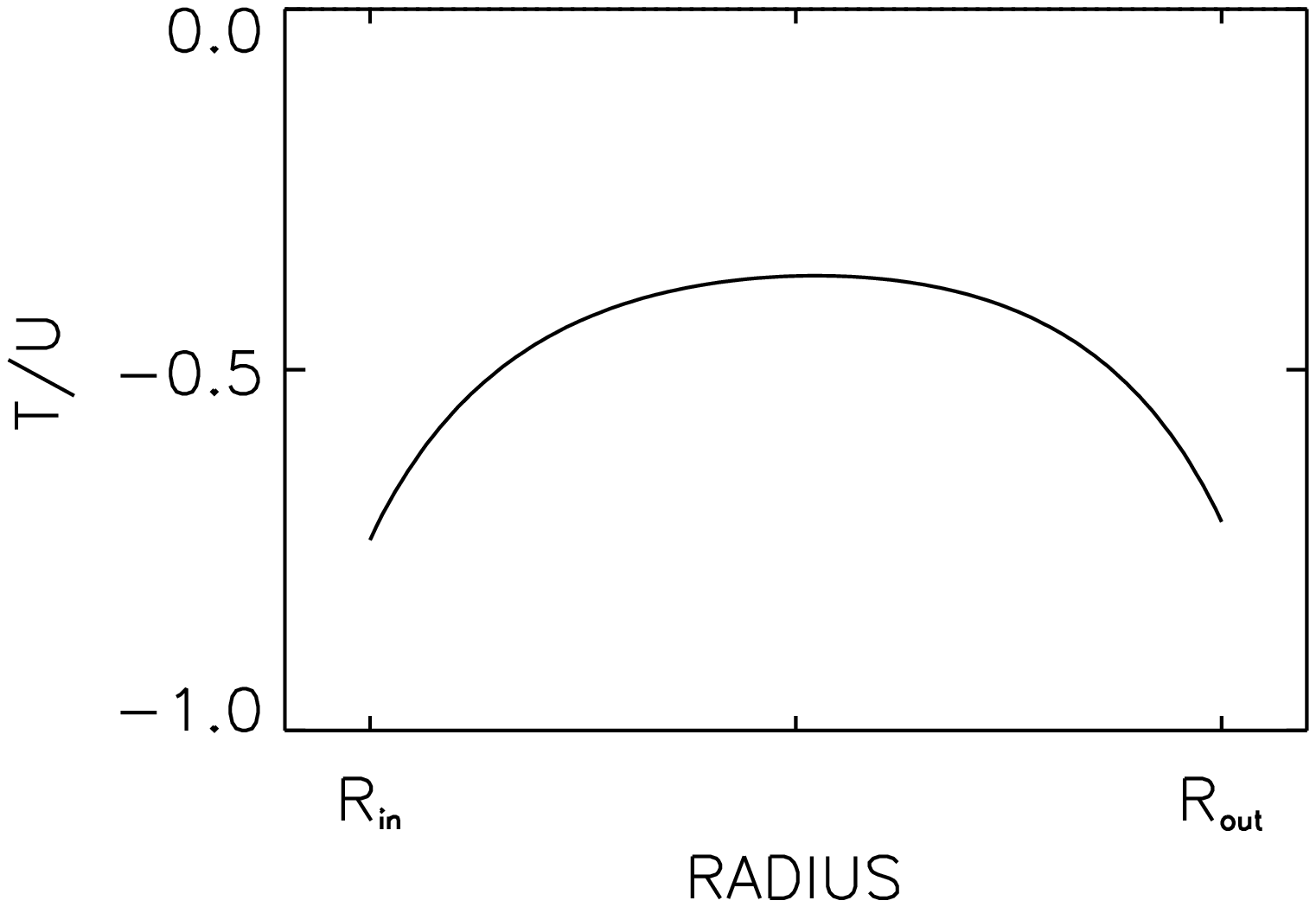}}
 \caption{The angular momentum transport (\ref{T}) for  subrotation ($\mu_\Om= 0.92$, top) and superrotation ($\mu_\Om= 1.07$, bottom). $\rm Re=500$, $\rm Pm=1$.   The torque is positive for subrotation and negative for superrotation. The values are normalized with the turbulence intensity $U=\sqrt{\langle u_R'^2 \rangle\langle u_\phi'^2 \rangle}$.  Due to the Maxwell stress they can exceed unity.}  
    \label{amt}
 \end{figure}

After  normalization the  expression (\ref{T}) reads
\beg
T_R\simeq \langle u_R' u_\phi'\rangle - {\rm Ha}^2 {\rm Pm} \langle B_R' B_\phi'\rangle.
\label{norm}
\ende
In Fig. \ref{amt} the angular momentum (\ref{T}) normalized with $\sqrt{\langle u_R'^2 u_\phi'^2 \rangle}$ is given. Without magnetic fields its absolute  value must be smaller than unity. The Maxwell stress, however, may produce higher values. We take from  Fig. \ref{amt} that in the linear theory the magnetic contribution is   surprisingly small.
 
 For simplicity we  only work here with ${\rm Pm}=1$.  The angular momentum transport vanishes for rigid rotation and it is indeed anticorrelated with $\nabla \Om$. The diffusion approximation (\ref{T1})  is thus possible. Its magnetic part is (only) of the same order than its kinetic part. 
\section{Wide gap} 
For $\hat\eta=0.05$ it is $\mu_B=1/\hat\eta=20$. The critical Hartmann number for $\rm Re=0$ is 0.31 for all Pm (see Fig. \ref{Ha}). The Rayleigh limit for centrifugal instability is $\mu_\Om=\hat\eta^2=0.0025$.
\subsection{Electric currents}
The technical possibilities are now discussed to realize this Hartmann number in the laboratory working with liquid metals.
Let $I_{\rm axis}$ be the axial current inside  the inner cylinder and $I_{\rm fluid}$ the axial current between the inner and the outer cylinder. The assumption $B=0$ in Eq. (\ref{basic})  provides 
\beg
\frac{I_{\rm fluid}}{R_{\rm out}^2-R_{\rm in}^2}= \frac{I_{\rm axis}}{R_{\rm in}^2}
\label{IfluIax}
\ende
as  the current density  is homogeneous. Hence,
\beg
\frac{I_{\rm fluid}}{I_{\rm axis}}=\frac{1-\hat\eta^2}{\hat\eta^2}= 399
\label{IfIax}
\ende
for $\hat\eta=0.05$. It  is also clear  that
\beg
B_{\rm in}=\frac{I_{\rm axis}}{5R_{\rm in}},
\label{Bin}
\ende
where $R$, $B$ and $I$ are measured  in cm, Gauss and Amp.
\begin{figure}[h]
   \includegraphics[width=8.5cm,height=6.0cm]{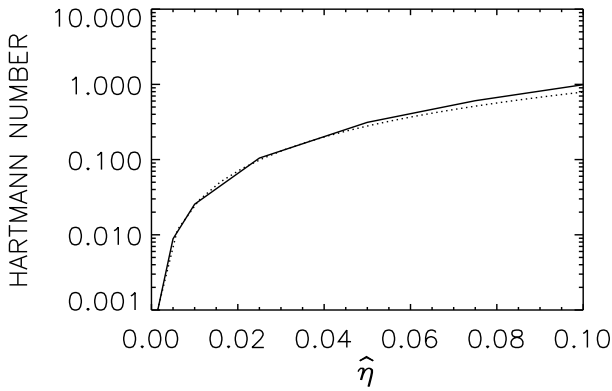}
  \caption{The same as in Fig. \ref{Ha1} but for $\hat\eta\to 0$. The dotted line represents the algebraic expression $25\cdot {\hat\eta}^{1.5}$.}  
    \label{Ha}
 \end{figure}
 It follows
\beg
I_{\rm axis}=\frac{5R_{\rm in} {\rm Ha}}{D} \sqrt{\mu_0 \rho \nu \eta}= 5 \sqrt{\frac{\hat\eta}{1-\hat\eta}} {\rm Ha} \sqrt{\mu_0 \rho \nu \eta}.
\label{Iaxis}
\ende
With the limit (\ref{Heta}) of Ha for small $\hat\eta$ one finds
\beg
I_{\rm fluid}= 140  \sqrt{1-\hat\eta} (1+\hat\eta)\sqrt{\mu_0 \rho \nu \eta} \ \ [{\rm Amp}] .
\label{current}
\ende
With the numerical values for $\hat\eta=0.05$ and  $\sqrt{\mu_0 \rho \nu \eta}=25.6$ for  the gallium-tin alloy used in the experiment PROMISE we find
\beg
I_{\rm fluid}=11.8 \ {\rm Ha}  \ \ [{\rm kAmp}]
\label{Ifluid}
\ende
for the  electric current through the gallium and 
\beg
I_{\rm axis}=29.5 \ {\rm Ha} \ \ [{\rm Amp}]
\label{Iaxis1}
\ende
for the current along the axis.
With ${\rm Ha}=0.31$  for $\hat\eta=0.05$ (see Fig. \ref{Ha}) the results are $I_{\rm fluid}=3.66$ kAmp and  $I_{\rm axis}=9.1$ Amp.

In the limit $\hat\eta\to 0$ the total current through the fluid conductor becomes 3.20 kAmp which is within the present-day technical possibilities. It should  thus be possible to realize the nonaxisymmetric current-driven TI in the MHD laboratory also with fluids of small magnetic Prandtl number, e.g. with sodium and  gallium.
\subsection{Uniform rotation}
\begin{figure}[h]
   \includegraphics[width=8.5cm,height=6.0cm]{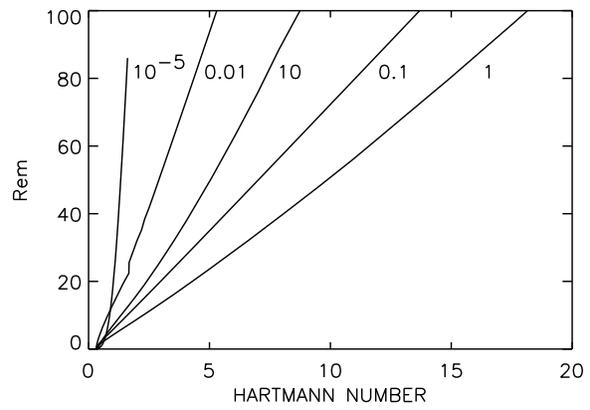}
  \caption{The same as in Fig. \ref{fig5a} but for $\hat\eta=0.05$. The curves are marked with the  magnetic Prandtl numbers. Again rotating fluids  with $\rm Pm=1$ undergo the  strongest stabilization.}  
    \label{rem2}
 \end{figure}

Figure \ref{rem2} gives for the wide-gap container the rotational quenching of the TI for various values of the magnetic Prandtl number Pm similar to  Fig. \ref{fig5a} for the narrow gap. Again the lines for marginal instability in the Rem-Ha plane are  straight lines. The line for $\rm Pm=1$ gives the ultimate stabilization by rigid rotation. A stronger stabilization does not exist. Hence, if ${\rm Rem}\lsim 6 \cdot {\rm Ha}$ then the fluid is always unstable (for rigid rotation). The rotational stabilization is much weaker for all $\rm Pm\neq 1$. It is in particular weak for very small Pm.  

\begin{figure}[h]
     \includegraphics[width=8.5cm]{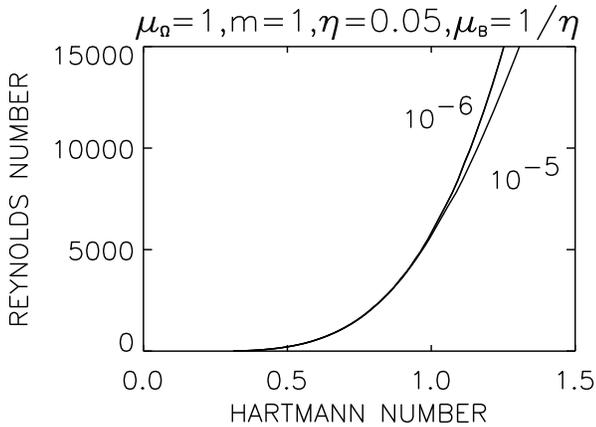}
     \caption{The  suppression of the TI in a wide-gap container by rigid rotation for   $\rm Pm = 10^{-5}$ and $\rm Pm = 10^{-6}$. The standard Reynolds number (\ref{re}) is given for experimental applications. $\hat\eta=0.05$,  $\mu_B=20$.}
    \label{f11}
 \end{figure}

The  question is whether also    the rotational stabilization of the TI can be probed in the laboratory. Therefore,   the rigidly  rotating wide-gap container is considered also for the very small magnetic Prandtl numbers of fluid metals. The magnetic Prandtl number of sodium is about $\rm Pm = 10^{-5}$, and for gallium it is about 
$\rm Pm = 10^{-6}$. Without rotation the critical Hartmann number is 0.31 for this container (independent of  Pm). With rotation the numerical results  are given in Fig. \ref{f11}. We find the rotational stabilization also existing  for fluids with their small Pm.  For not too fast    rotation the differences of the resulting critical Ha are very small  for both the fluid conductors   so that  experiments with gallium are also  possible.  For a (small) Reynolds number of order 1000 the marginal-stable magnetic field is about two times higher than for $\rm Re=0$. It should thus not be too  complicated to find the basic effect for the rotational suppression of TI -- which proves to  be important  both for the rapid-rotating  hot MS stars and also for neutron stars  --  in the MHD laboratory.

The curves in Fig.  \ref{f11}  do not become more steep for $\rm Pm\to 0$. There is no visible difference between the curves for ${\rm Pm}=10^{-6}$ and ${\rm Pm}=0$.

 \section{Conclusions}
 In this paper  the interplay of Tayler instability and rotation for incompressible fluids of uniform density filling the gap between 
 the cylinders of a Taylor-Couette container is considered. The toroidal field is the result of an electric 
 current of homogeneous density. It is shown that for zero rotation the critical magnetic field amplitudes 
 for marginal instability 
 does not depend on the magnetic Prandtl number. The critical magnetic field  strongly depends on the gap width. It is very high for small gaps and it this rather low for wide gaps. For small enough inner radius $R_{\rm in}$ the critical Hartmann number (of the inner field) runs as $R_{\rm in}^{1.5}$. The resulting electric currents necessary for TI with $m=1 $ are   3.66 kAmp if the material is the same gallium-tin alloy as used in the experiment PROMISE. Such currents can easily be produced in the laboratory.

For a narrow gap with   $\hat\eta=0.95$ the rotational quenching of the TI is  studied in detail. Figure \ref{fig5a} displays the rotational stabilization   for various magnetic Prandtl numbers. For the  normalization of  the basic rotation a `mixed' Reynolds number (\ref{rem}) is used in which  -- as in the Hartmann number -- the viscosities $\nu $ and $\mu$ are symmetric. The ratio of this  Reynolds number Rem and the Hartmann number is basically free of any diffusivity. For fast enough rotation just this ratio describes the rotational quenching of TI for various Pm. In this representation the results for Pm between 0.01 and 10 are rather simple. The most effective stabilization of TI happens for $\rm Pm=1$. It is weaker for both smaller Pm and  higher Pm. This is an unexpected result which may warn that many numerical simulations with $\rm Pm\simeq 1$ could  overlook the nonaxisymmetric instability of strong toroidal fields. 

Also the inclusion of differential rotation leads to surprising results. For $\rm Pm=1$ the Fig. \ref{f3}  presents the basic differences for rotation laws with different signs of  ${\rm d}\Om/{\rm d} R$. While superrotation always stabilizes the  magnetic field, there is a dramatic destabilization phenomenon  by subrotation for  medium rotation rates. This is the effect announced  with   Eq. (\ref{ach}) by Acheson (1978). For  slow and fast rotators (old and young stars with outer convection zones) with  negative shear the  toroidal magnetic fields   have a very different stability behavior. Slowly rotating (old) stars with small Rem can accumulate much stronger magnetic fields than young stars with their larger Rem. However,  if the  rotation is too strong then  the nonaxisymmetric instabilities are  more and more destroyed by the very strong shear  (see Fig. \ref{f6}). 
   
Our calculations  demonstrate  that rigid rotation always stabilizes the magnetic field against the nonaxisymmetric TI. We have also  shown  that  this rotational stabilization should  be observable in the laboratory. Figure \ref{f11} provides the result that the critical Hartmann number in a wide-gap container of  $\hat\eta=0.05$ can be increased by a factor of two by a rigid rotation of Reynolds number 1000. For the gallium-tin alloy with   its  molecular viscosity of $3.4\cdot 10^{-3}$ cm$^2$/s this Reynolds number is reached for a rotation frequency of about $11.4/R_{\rm out}^2$ Hz with $R_{\rm out}$  in cm. The rotation  frequency of  0.11 Hz for $R_{\rm out}\simeq 10$ cm is rather small. For the same container one needs  the  electric current of  7.32 kAmp through the  gallium-tin alloy to realize the Tayler instability  in the rotating fluid conductor.


\end{document}